\documentclass[pre,aps,twocolumn]{revtex4-1}

\usepackage[english]{babel}
\usepackage{graphicx}
\usepackage{amsfonts}
\usepackage{amsmath}
\usepackage{color}

\begin{document}

\title{Dynamical complexity measure to distinguish organized from disorganized dynamics}

\author{Christophe Letellier}
\affiliation{Rouen Normandie University -- CORIA, Avenue de l'Universit\'e, 
F-76800 Saint-Etienne du Rouvray, France}

\author{I. Leyva}
\affiliation{Complex Systems Group \& GISC,  Universidad Rey Juan Carlos, 28933 M\'ostoles, Madrid, Spain}

\affiliation{Center for Biomedical Technology, Universidad Polit\'ecnica
de Madrid, 28223 Pozuelo de Alarc\'on, Madrid, Spain}

\author{I. Sendi\~na-Nadal}
\affiliation{Complex Systems Group \& GISC,  Universidad Rey Juan Carlos, 28933 M\'ostoles, Madrid, Spain}

\affiliation{Center for Biomedical Technology, Universidad Polit\'ecnica
de Madrid, 28223 Pozuelo de Alarc\'on, Madrid, Spain}

\date{\today}

\begin{abstract}
We propose a metric to characterize the complex behavior of a dynamical 
system and to distinguish between organized and disorganized complexity. The 
approach combines two quantities that separately assess the degree of 
unpredictability of the dynamics and the lack of describability of the 
structure in the Poincar\'e plane constructed from a given time series. As for 
the former, we use the permutation entropy $S_{\rm p}$, while for the later, we 
introduce an indicator, the {\em structurality} $\Delta$, which accounts for 
the fraction of visited points in the Poincar\'e plane. The complexity measure  
thus defined  as the sum of those two components is validated by classifying in 
the ($S_{\rm p}$,$\Delta$) space the complexity of several benchmark 
dissipative and conservative dynamical systems. As an application, we show how 
the metric can be used as a powerful biomarker for different cardiac 
pathologies and to distinguish the dynamical complexity of two electrochemical dissolutions.
\end{abstract}

\maketitle

\section{Introduction}

Recent decades have witnessed a considerable growth of the science of 
complexity, devoted to understanding the collective behavior of systems, 
regardless of their physical, biological, or social 
nature\cite{Ada00,Per05,Reg06,Sab09,Fla11,Aqu17,Poz17}. Much of the research 
has focused on defining multiple metrics to classify and quantify complex 
dynamics involving many variables \cite{Llo01}. First attempts were made by 
extending information theory  to dynamical systems \cite{Kol58,Sin59} and 
adapting Shannon's entropy \cite{Sha48} to statistically estimate the apparent 
randomness present in deterministic chaotic dynamics. As long as the Jaynes' 
principle of maximum entropy \cite{Jay57} is properly applied 
\cite{Ski84,Ber96,Lar92,Fre15}, the Shannon's entropy informs about the rate at 
which information is produced and, consequently, it is a measure of a system's 
predictability. However, as Weaver posited \cite{Wea48}, there are two classes 
of complexity: disorganized and organized. We will refer here to organized 
(disorganized) dynamics when its Poincar\'e section is describable 
(indescribable). While the former can be tackled using the methods of 
statistical mechanics and probability theory, the latter cannot be fully 
understood using statistics alone as it involves considerable large number of 
variables that are interrelated.

Entropic metrics work well as indicators of the level of unpredictability and 
randomness, but fail to correctly capture the existence of inter-dependencies 
or structure among the system's components \cite{Hub86}. This drawback is mostly due to the commonly accepted standard that both maximally random and perfectly ordered systems do not exhibit any degree of structural organization \cite{Hub86} and, therefore, a measure quantifying their degree of complexity should be minimal 
\cite{Fel98} without any distinction between the two cases. Among several approaches proposed for detecting an underlying structure, are those called {\it statistical} complexity measures, which 
account for the graph complexity of the representation of symbolic sequences 
as trees \cite{Hub86,Cru89}, or in terms of disturbance from the equiprobable 
distribution of the accessible  states of the system \cite{Lop95,Mar03}. 
However, these strategies, although using combinations of different 
indicators, rely on the same background information: the probability of 
different symbolic sequences. Thus, for instance, they are not able to 
distinguish a fully developed chaos from a stochastic dynamics as in both cases 
the probabilities are uniformly distributed.

Here we provide a {\it dynamical} complexity measure $C_{\rm D}$ which is able 
to rank  both the degree of unpredictability and indescribability of a 
structure present in a process. It combines the Shannon entropy as indicator 
of the unpredictability with the density of points in the Poincar\'e plane as 
an alternative metric of organization. Our measure is designed to be zero for a 
fully predictable and perfectly ordered dynamics, one for a nonpredictable but 
organized dynamics and a value of two for a nonpredictable disorganized one. 
We illustrate the capabilities of our measure introduced in Sec. \ref{dyco} 
in several dissipative and conservative dynamical systems whose complex 
behavior can be accessed by tuning a system's parameter (Sec. \ref{besy}). 
Finally, in Sec. \ref{data}, we show how it can be used to compare the complex 
dynamics of two electrochemical dissolutions and as  a biomarker for different 
cardiac diseases directly obtained from electrocardiograms (ECGs).

\section{Dynamical complexity}
\label{dyco}

\begin{figure}
  \centering
  \includegraphics[width=0.47\textwidth]{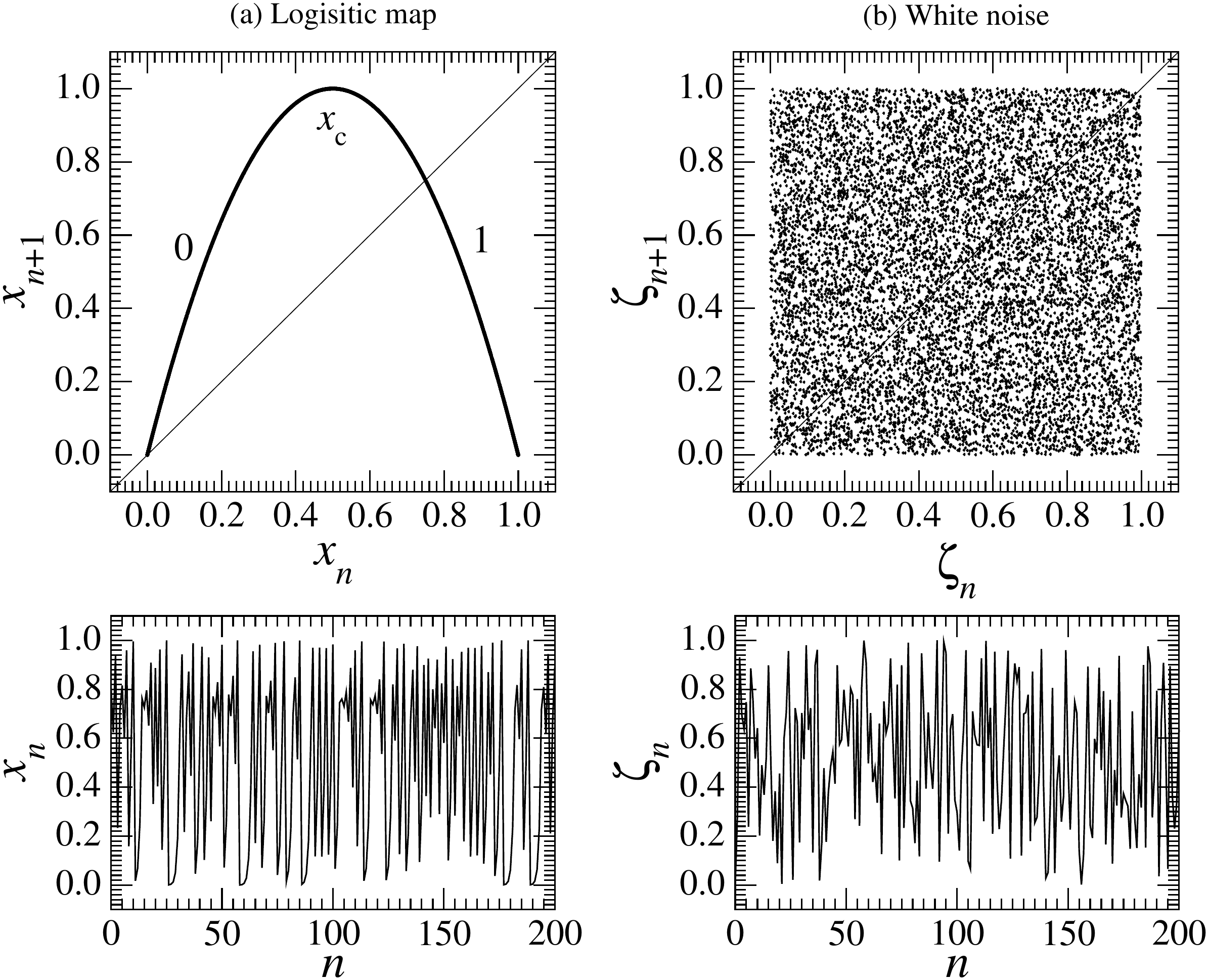} \\[-0.3cm]
  \caption{(Upper row) First-return maps for (a) logistic map with $\mu = 4$ 
and (b) uniform white noise with the same amplitude. (Bottom row) 
Characteristic time series for the same both examples. For the logistic map, 
the first return map shows the generating partition $x_{\rm c} =\frac{1}{2}$ and the 
corresponding symbolic dynamics.}
  \label{char}
\end{figure}

Let us start by showing how the family of the extensively used entropy-based
statistical complexity measures $C_{\rm S}$, first introduced by L\'opez-Ruiz 
{\it et al.} \cite{Lop95} and later on improved by Martin {\it et al} 
\cite{Mar03}, 
fails to detect the organized complexity underlying the paradigmatic logistic 
map. $ C_{\rm S}$ is usually defined as $C_{\rm S}=Q H$, where $Q$ stands for 
the so-called disequilibrium, which quantifies how far is the probability 
distribution of a given process from the uniform one, and $H$ is the 
corresponding normalized Shannon entropy. Such a factorization ensures that 
$C_{\rm S}$ vanishes for perfect order ($H=0$) and maximal randomness ($Q=0$), 
and it is expected to capture a wide range of complex behaviors in between. 
However, when we compare a uniform white noise $\zeta_n \in [0;1]$ with the 
logistic map 
\begin{equation}
  x_{n+1} = \mu x_n (1 - x_n)
  \label{logistic}
\end{equation}
whose behavior $x_n \in [0;1]$ depends on the parameter $\mu$, we find a clear 
example where 
the statistical complexity measure $C_{\rm S}$ is not performing properly, in 
particular when the logistic map exhibits fully developed chaotic behavior for 
$\mu=4$. Figure \ref{char} shows an illustrative characterization of the two 
dynamics by means of first-return maps (top panels) and time series (bottom 
panels). The time series of both dynamics look quite alike, actually 
characterized by an almost flat histogram for the symbolic sequences 
\cite{Let08}, which yields to null values of $C_{\rm S}$. However, their 
respective first-return maps reveal a well-defined underlying structure for the 
chaotic dynamics [top panel in Fig.~\ref{char}(a)] while the noise fills the 
whole available state space [top panel in Fig.~\ref{char}(b)] with no signs of 
dynamical organization. 

Therefore, we need a marker capable of discriminating the presence of a 
structured dynamics, easily describable. The Shannon entropy already 
quantifies how the structure --- when it exists --- is visited in the state 
space. It must be complemented by a second marker capable of measuring how the 
structure fills the state space, based on the principle that the more 
structured the attractor is, the smaller the volume it occupies.  If we choose 
the Poincar\'e section as a more reliable source to compute the entropy than 
the time series itself \cite{Let06b}, then the argument can be reformulated by 
stating that the smaller the fraction of boxes visited by the Poincar\'e 
section, the more constrained and organized the dynamics. In general, we will 
be working in a two-dimensional projection of the Poincar\'e section 
independently of the system's dimension. For the strongly dissipative systems 
we investigate, the Poincar\'e section can be safely reconstructed from a 
single variable by using delay coordinates and we will therefore use the 
first-return map to obtain it.

To assess how a dynamics is structured in the state space, we 
introduce the {\it structurality} $\Delta$ index, which accounts for the 
fraction of visited boxes from a pixelation of the Poincar\'e section into 
$N_{\rm b} \times N_{\rm b}$ boxes, that is, 
\begin{equation}
  \Delta = {\sum^{N_{\rm b}}_{i,j=1} \displaystyle 
     \frac{v_{ij}}{N_{\rm b}^{2}}  } \in [0;1] \, ,
\end{equation} 
where $v_{ij}=1$ if the box $(i,j)$ contains at least one crossing point and 
$v_{ij}=0$ otherwise. To implement this definition we need 
first to determine the frame in which the Poincar\'e section is investigated. The easiest way is to construct a domain from the visited range, that is, the minimum and maximum values along the two axis of the Poincar\'e plane. We will refer to this frame as the {\it renormalized} frame. In some cases, as when a bifurcation diagram is investigated, 
it is more efficient to use a fixed frame corresponding to the most developed Poincar\'e section: we will designate it as a 
{\it relative} frame. Let $[m_i;M_i]$ be the range of the $i$ axis of the chosen frame. Then we need to specify the pixelation, that is, the number $N_{\rm b}^{2}$  of boxes, and their length $l_{\rm b}$. The former can be estimated as a function of the number $N_{\rm p}$ of points in the Poincar\'e section as 

\begin{equation}
N_{\rm b} \gtrsim 10 \, \log_{10}N_{\rm p} \, .  
\end{equation}
However, as long as $N_{\rm p}$ is large enough for properly sampling the 
dynamics, the dependence of $\Delta$ on the pixelation settings is not 
critical, as we show in the Appendix. The length of each box is 
determined according to 
\begin{equation}
  \label{domps}
  l_{\rm b} = \frac{\lVert M_i - m_i \rVert}{N_{\rm b}}
  \, , 
\end{equation}
A lack of describability of the structure of the dynamics 
will be manifested by a large value of $\Delta$, while a well ordered dynamics 
like a period-one behavior will have $\Delta = 1/N_{\rm b}^{2} \sim 0$. 
See the Appendix for a more elaborated discussion on this issue.

\begin{figure}[t]
  \centering
  \includegraphics[width=0.45\textwidth]{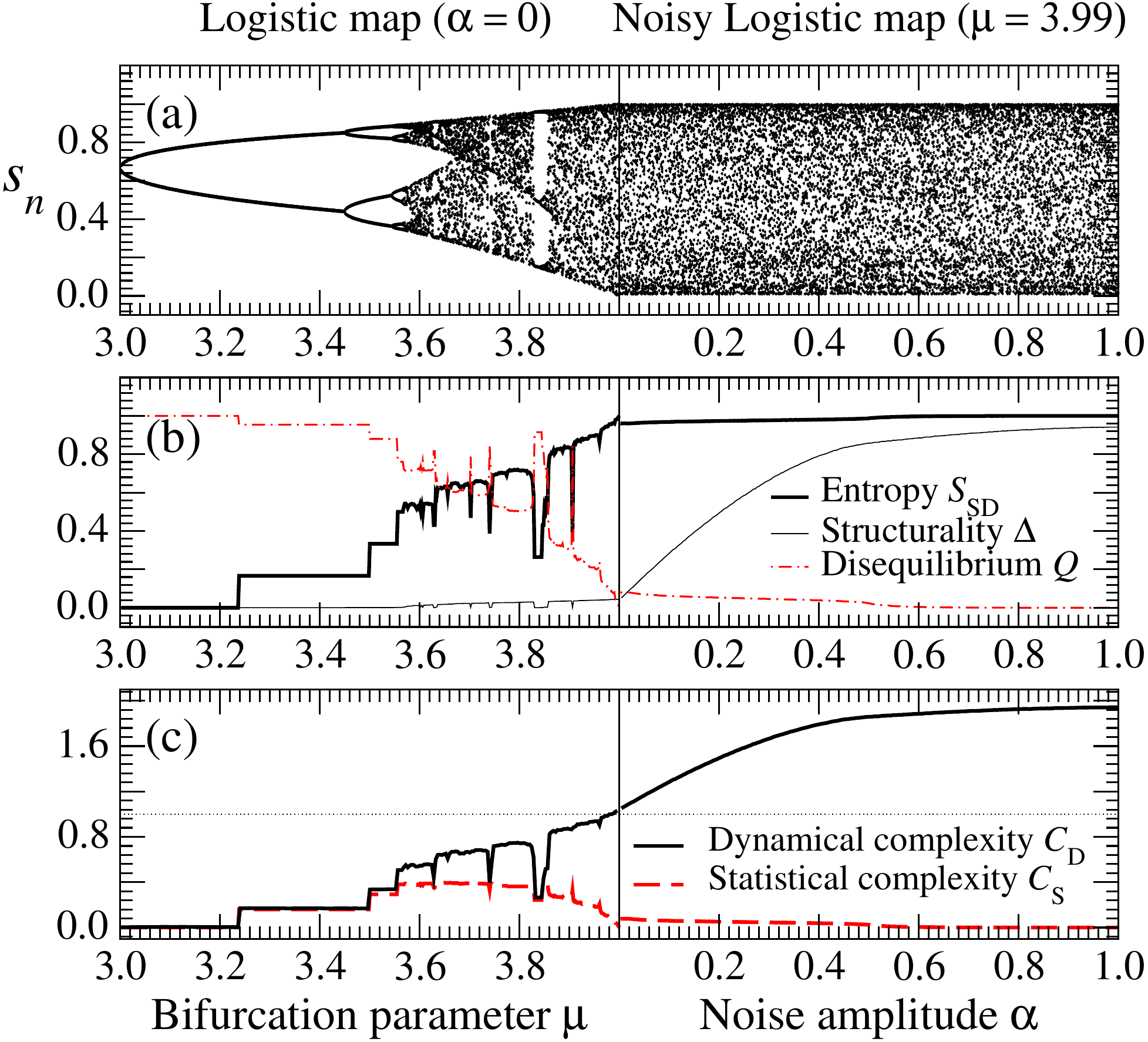} \\[-0.2cm]
  \caption{(a) Bifurcation diagram of the logistic map vs $\mu$ for $\alpha=0$ 
(left half) and vs $\alpha$ for $\mu=4$ as it is progressively 
replaced with noise (right half) according to Eq.~(\ref{aggreg}). (b) Shannon 
entropy based on the symbolic dynamics $S_{\rm SD}$ (black thick line), relative
structurality $\Delta$ (black thin line), and disequilibrium $Q$ (red 
dash-dotted line). (c) Dynamical complexity $C_{\rm D}=S_{\rm SD}+\Delta$ and 
statistical complexity $C_{\rm s} = S_{\rm SD}\cdot Q$. $N_{\rm p} = 180,000$. }
\label{logis}
\end{figure}

At this point, we combine into a single dynamical complexity measure 
$C_{\rm D}$ the two components characterizing the lack of  predictability, the 
Shannon entropy $S_{\rm p}$, and the lack of describability, the structurality 
$\Delta$, as
\begin{equation}
  C_{\rm D} = S_{\rm p} + \Delta \, , \label{Cd}
\end{equation}
where $S_{\rm p}$ is by default the permutation entropy \cite{Ban02}, unless 
stated otherwise. 

We evaluate now how $C_{\rm D}$ performs by first discussing its application to the logistic map $x_n$ {dynamically} coupled to a white { uniform} noise $\zeta_n$ as 
\begin{equation}
  \label{aggreg}
  s_n = (1 - \alpha) x_n + \alpha \zeta_n,
\end{equation}
where  $x_n, \; \zeta_n \in [0;1]$ and $\alpha$ is a parameter ranging between 
$0$ ($s_n=x_n$, logistic map) and $1$ ($s_n=\zeta_n$, white noise). Here the 
Shannon entropy $S_{\rm SD}$ is based on the symbolic dynamics produced by the 
generating topological partition \cite{Let06b} $\sigma_i =0$ if 
$s_n<\frac{1}{2}$ and $\sigma_i=1$ otherwise, which satisfies the 
maximum-entropy principle for the logistic map \cite{Jay57}. The dynamics is 
thus reduced to {sequences} of $0$'s and $1$'s of length 6 whose frequencies of 
occurrence are calculated over $N_{\rm p} = 180,000$.

The left half of Fig.~\ref{logis}(a) shows the bifurcation diagram of the noise 
free ($\alpha=0$) logistic map for $3 \le \mu \le 4$, while in the right half 
$\mu$ is kept constant to $3.99$ and $0 \le \alpha \le 1$, giving rise to a 
fully developed chaos increasingly contaminated by uniform noise of growing 
amplitude. In Fig.~\ref{logis}(b) we plot the Shannon entropy $S_{\rm SD}$,
the structurality $\Delta$, and the generalized Jensen-Shannon disequilibrium 
$Q$ (red dash-dotted line) as proposed in Ref. \cite{Mar03}. $S_{\rm SD}$ 
initially increases in a stepwise form as the logistic map undergoes the 
period-doubling bifurcation up to a point where a much richer structure arises 
characterized by chaotic behavior intermingled with periodic windows in which 
the entropy drops accordingly to the periodicity. For $\mu = 4$, the logistic 
map is fully chaotic, all its symbolic states are equally likely and, 
therefore, $S_{\rm SD}=1$, reaching the maximum lack of predictability of the 
system, which keeps bounded independently of the added noise intensity. 
However, the structurality $\Delta$ 
informs us about how organized is the state space: the Poincar\'e section of 
the noise-free logistic map changes from an isolated point for the period-1 
cycle ($\mu=3$) to a smooth unimodal map in the unit square for the fully 
developed chaos [$\mu=4$, see Fig.~\ref{char}(a)], yielding in every case to a 
very well structured dynamics with $\Delta \ll 1$, but still capturing the 
different degrees of chaotic behavior. When noise is added to the $\mu = 4$ 
case [right half in Fig.~\ref{logis}(b)], $\Delta$ increases monotonously until 
it saturates when the lack of structure fills up completely the unit square and 
$\Delta \sim 1$. Note that, while the entropy barely changes, the relative 
structurality 
is clearly differentiating the increasing degree of disorganization of the 
dynamics. Regarding the disequilibrium, it behaves as expected as it is again a 
function of the probability of the different symbolic sequences: it is maximum 
for period-1 oscillations and vanishes indistinguishably for fully chaotic and 
stochastic dynamics. Finally, in Fig. \ref{logis}(c) dynamical and statistical 
complexities are depicted together. While $C_{\rm S}$ exhibits the well-known 
behavior exclusively driven by the information contained in the entropy, 
$C_{\rm D}$ is able to discriminate between the different dynamical behaviors 
in increasing order of complexity, assigning the maximum value $C_{\rm D}=2$ to 
the fully unpredictable and indescribable dynamics featured by a white noise, 
the minimum value $C_{\rm D}=0$ to a fully predictable and describable periodic 
motion, and a value close to $1$ for a yet unpredictable but structured 
(describable) chaotic behavior.  

\section{Benchmark systems}
\label{besy}

\begin{figure}
  \centering 
  \includegraphics[width=0.36\textwidth]{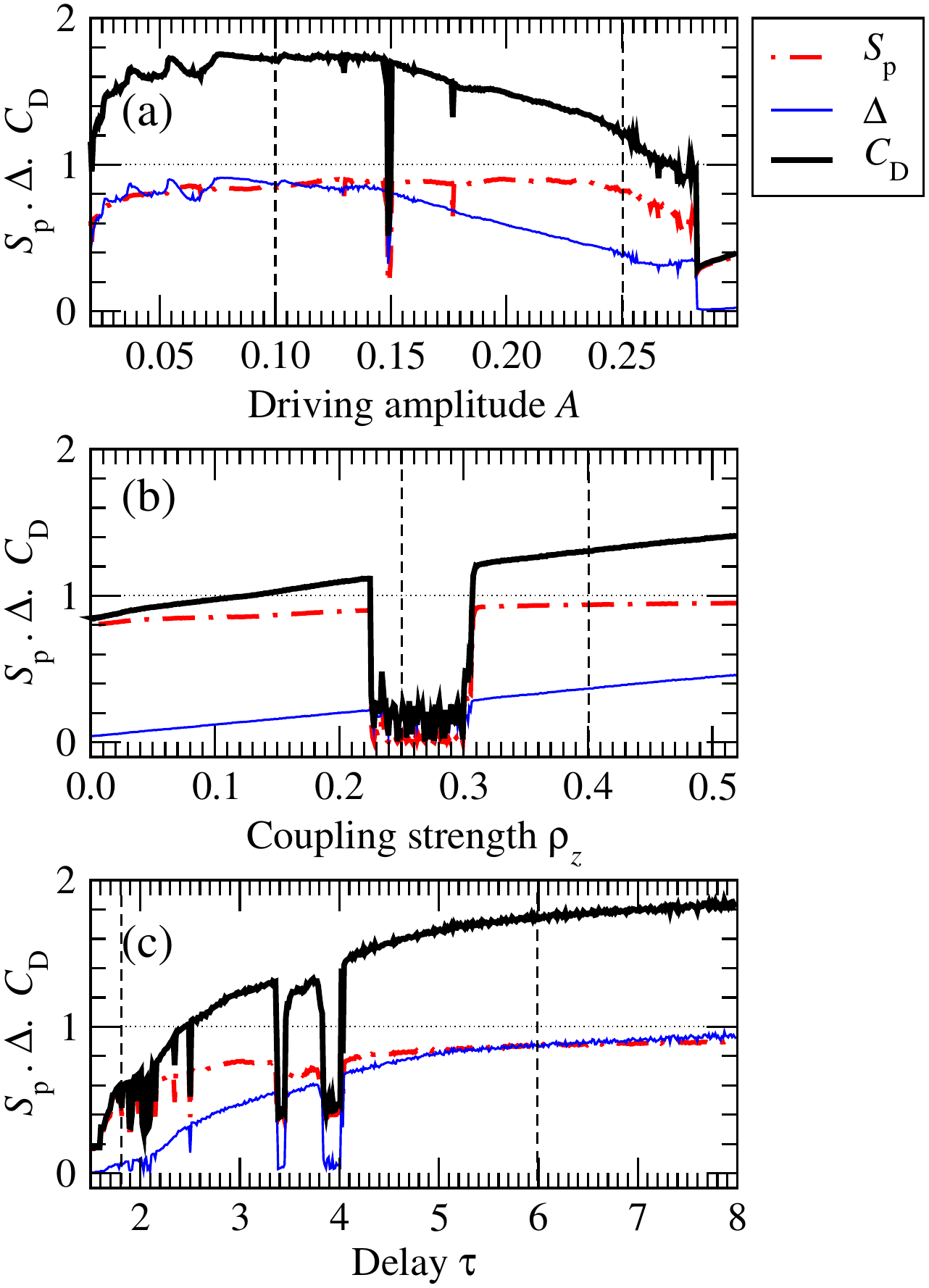} \\[-0.4cm]
  \caption{Evolution of the permutation entropy $S_{\rm p}$, relative 
structurality $\Delta$, and dynamical complexity $C_{\rm D}$ versus a 
bifurcation parameter for (a) the symmetrized double-gyre system, (b) a dyad 
of R\"ossler systems and (c) the delay-differential Mackey-Glass equation. 
Legend in (a) applies to all panels. Vertical dashed lines mark some 
parameter values whose dynamical regimes will be later analyzed more in 
detail.
  \label{bifcomp}}
\end{figure}

So far, we have illustrated the definition of our dynamical complexity measure 
$C_{\rm D}$ using a simple nonlinear map. To test it in a more 
general context of flows, we have chosen three dynamical systems whose state 
space range from finite to infinite dimensional and whose dynamical behavior 
can be varied with a parameter: (i) a 4D double-gyre conservative system 
\cite{Cha19}, (ii) two coupled R\"ossler oscillators \cite{Ros78}, and (iii) 
the time delay-differential Mackey-Glass model \cite{Far82}. 
{Let us show the results before introducing each system and discuss 
the connection between the complexity measure and the dynamical interpretation.}
Figure \ref{bifcomp} shows the evolution of the dynamical complexity and its 
two components, $S_{\rm p}$ and $\Delta$ versus the corresponding bifurcation 
parameter. 

\begin{figure}
  \centering
  \includegraphics[width=0.35\textwidth]{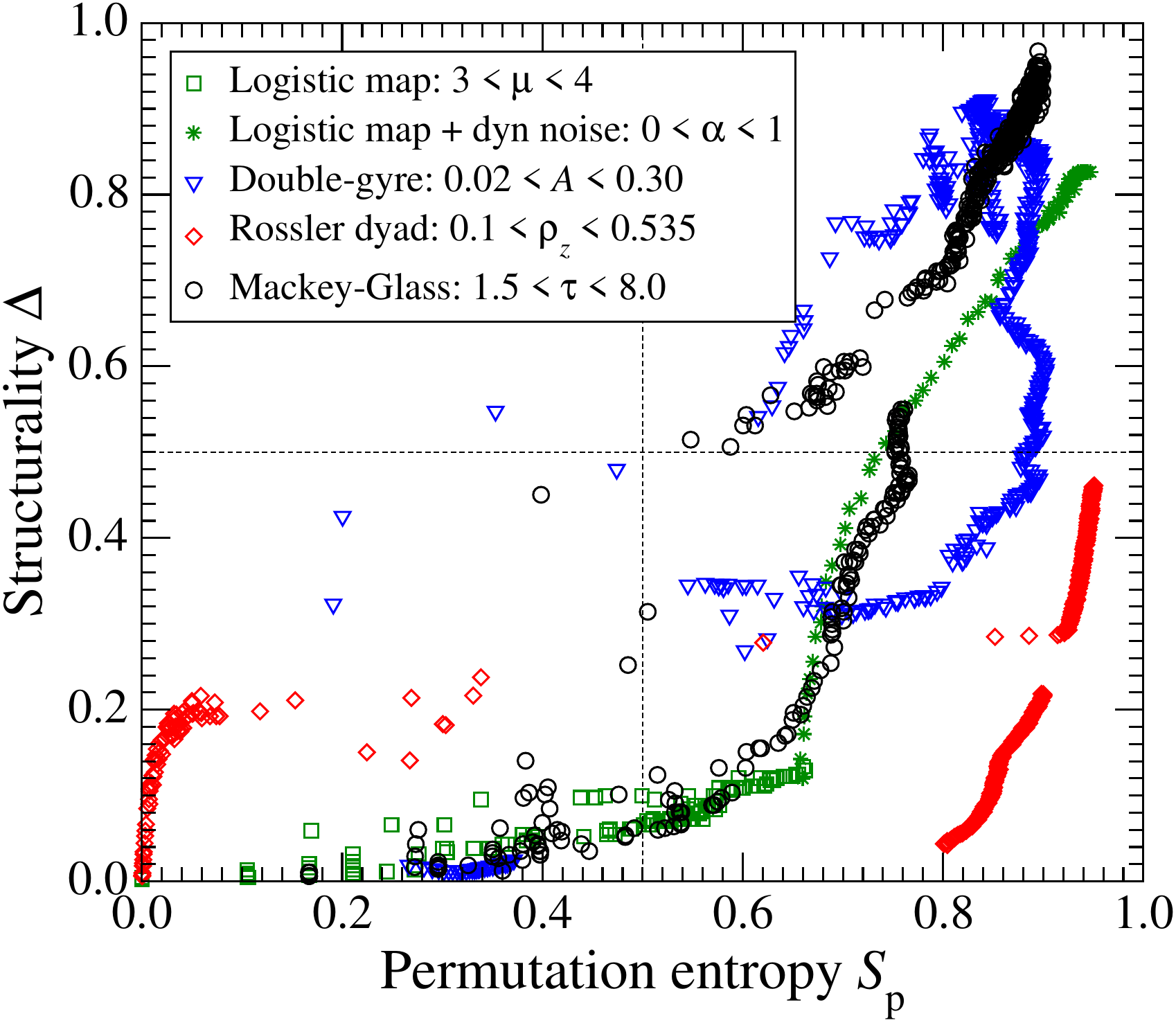} \\[-0.3cm]
  \caption{Permutation entropy $S_{\rm p}$ versus relative structurality 
$\Delta$ for the same systems and parameters as in Figs.~\ref{logis} and 
\ref{bifcomp}. }
 \label{ShDmap}
\end{figure}

The results of Figs.~\ref{logis} and ~\ref{bifcomp} are further explored in 
Fig.~\ref{ShDmap} plotting the evolution of $\Delta$ versus $S_{\rm p}$ of each 
system along their respective parameters. Remarkably, in this last 
representation the different dynamical 
regimes are almost exclusively distributed in three regions according to our 
complexity descriptor: lower-left region ($S_{\rm p}, \Delta<0.5$), 
corresponding to a very structured and predictable behavior; lower-right region 
($S_{\rm p} >0.5$, $\Delta<0.5$), structured but unpredictable behavior; and 
upper-right region ($S_{\rm p}$, $\Delta>0.5$) comprising unpredictable and 
indescribable dynamics. 

\subsection{Symmetrized double-gyre system}

{Conservative systems can produce chaotic as well as regular (periodic or
quasi-periodic) behavior. According to the KAM theorem, only a finite fraction 
of the state space is visited by regular solutions which can be viewed as
``islands'' in a chaotic sea \cite{Hen64,Lic92}. Typically, this chaotic sea is 
very difficult to describe and we expect to have a complexity measure 
${\cal C}_{\rm D}$ close to 2 \cite{Cou00}. As an example of this class of 
systems, we used a recently introduced simplified model for the driven 
double-gyre, a typical phenomenon observed in the large-scale ocean circulation 
\cite{Cou00,Sha05,Cha19}:}
\begin{equation}
  \label{DGsym}
      \begin{array}{l}
      \dot{x} = A \pi \sin [\pi (u x^2 +x-u)] \sin (\pi y) \\[0.1cm]
      \dot{y} = A \pi \cos [\pi (u x^2 +x-u)] \cos (\pi y) \\[0.1cm]
      \dot{u} = v \\[0.1cm]
      \dot{v} = - \omega^2  u \, . 
    \end{array}
\end{equation}
{
This is a four-dimensional conservative system where $x$ and $y$ are the variables spanning the physical space, $u$ and $v$ are variables related to the velocity field, and $A$ is the amplitude of a periodic forcing applied to the velocity field. The corresponding complexity markers are  shown in Fig.~\ref{bifcomp}(a) as a function of the driving amplitude $A$. 
}

\begin{figure}[ht]
  \centering
    \includegraphics[width=0.45\textwidth]{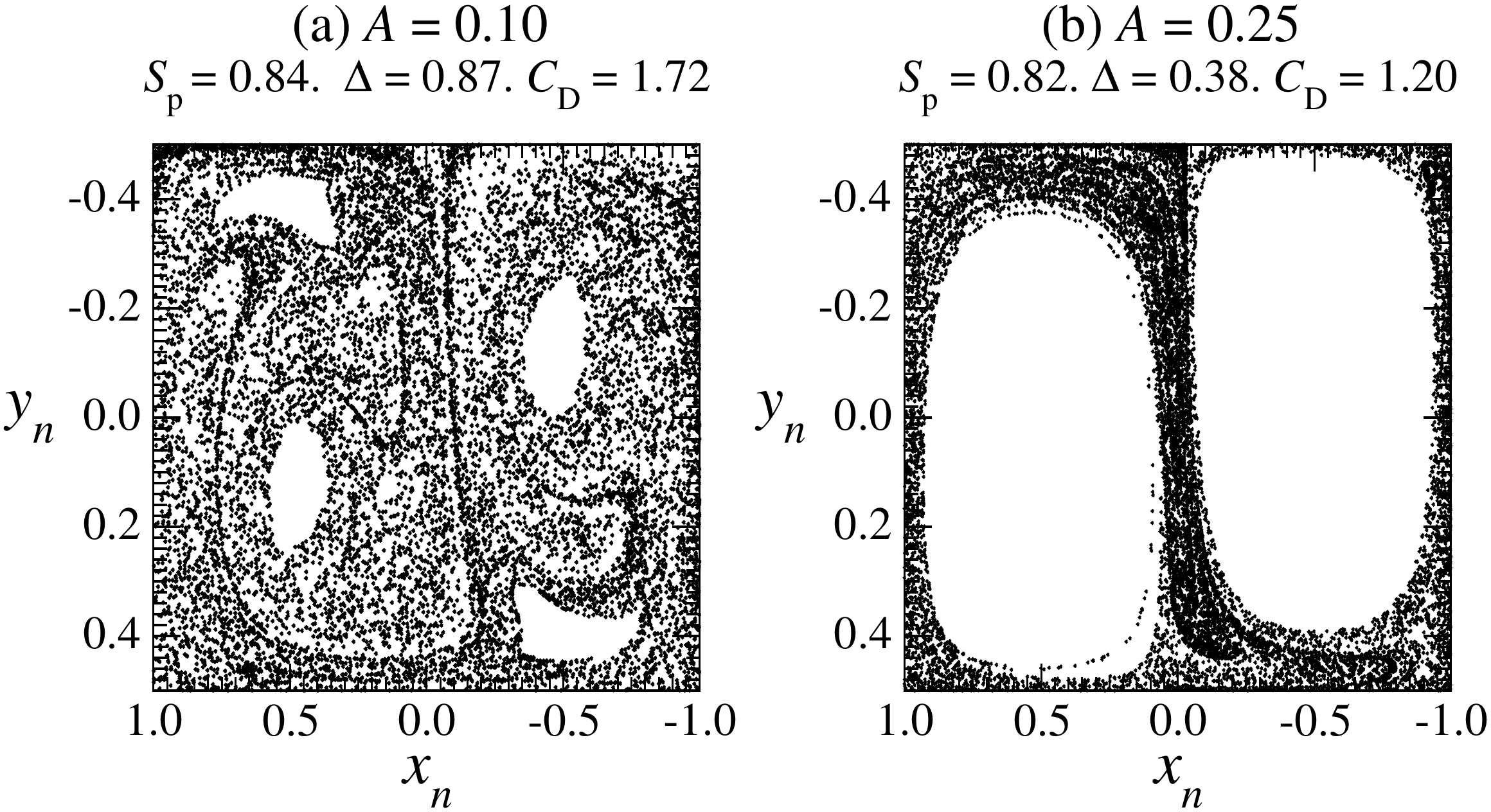} \\[-0.2cm]
  \caption{Poincar\'e section of the symmetrized driven double-gyre system
(\ref{DGsym}) for various initial conditions.  {The corresponding $A$ values are marked with vertical dashed lines in Fig.~\ref{bifcomp}(a).} Other parameter values:
$\omega = \frac{\pi}{5}$.}
  \label{doumap}
\end{figure}

{
The solutions produced by this system are investigated using the Poincar\'e
section
}
\begin{equation}
  {\cal P}_{\rm dg} \equiv 
  \left\{
    (x_n, y_n) \in \mathbb{R}^2 ~|~ u_n = 0, v_n > 0 
  \right\} \, . 
\end{equation}
{
The main difference between a strongly dissipative system and a conservative
one is that, in the former case, the Poincar\'e section is a uni-dimensional
curve while, in the latter case it is at least a two-dimensional structure: As
a consequence, the symmetrized driven double-gyre system must be investigated
in a two-dimensional Poincar\'e section to correctly understand the
organization of the chaotic sea around the regular islands. This is exhibited
by the Poincar\'e section $\{x_n, y_n\}$ in Fig.\ \ref{doumap}(a) where four 
large 
islands are observed. Since we are here concerned by the dynamical complexity
computed for the chaotic sea, we selected initial conditions in the middle of
the Poincar\'e section, a neighborhood which belongs to the chaotic sea (when
it exists) for most of the parameter values used in this work.}

Computations are performed with $N_{\rm p} = 80,000$ and a relative 
frame $[-1;+1] \times [-0.5;+0.5]$. Notice that all chaotic sea would induce a 
renormalized frame equal to this relative frame. As shown in 
Fig.~\ref{bifcomp}(a), this system presents high values of the Shannon entropy 
($S_{\rm p}>0.7$) and structurality ($\Delta \geq 0.4$), revealing that the map 
is not 1D. The size of the regular islands strongly depends on the $A$-value: 
for instance, when $A$ increases beyond $0.1488$, the size of these islands 
grows with a consistent decay in the complexity, therefore reducing the visited 
region which results in a reduction of $\Delta$. {This is well 
exemplified in Fig. \ref{doumap}(b), where the Poincar\'e section for 
$A = 0.25$ shows two very large regular islands}.
In fact, the structurality quantifies the mixing properties of this flow: the 
larger $\Delta$, the greater the mixing. 

\subsection{A dyad of R\"ossler systems}

The next example in Fig.~\ref{bifcomp}(b) is a 6D system of two slightly 
mismatched chaotic R\"ossler oscillators --- in a non phase-coherent regime 
--- diffusively coupled through the $z$ variable. The equations are:
\begin{equation}
  \label{ros76}
    \begin{array}{l}
      \dot{x}_1 = -y_1 -z_1 \\[0.1cm] 
      \dot{y}_1 = x_1 + a_1 y_1 \\[0.1cm]
      \dot{z}_1 = b + z_1 (x_1 -c ) + \rho_z (z_2 - z_1) \\[0.1cm]
      \dot{x}_2 = -y_2 -z_2 \\[0.1cm] 
      \dot{y}_2 = x_2 + a_2 y_2 \\[0.1cm]
      \dot{z}_2 = b + z_2 (x_2 -c ) + \rho_z (z_1 - z_2) 
    \end{array}
\end{equation}
{
with $a_1 = 0.492$, $a_2 = 0.480$, $b=2$, $c=4$ and initial conditions 
$x_1 = y_1 = z_1 = y_2 = z_2 = 0.2$, and $x_2=0.4$. The coupling through 
variable $z$ makes this system a class {\sc iii} system \cite{Boc06}, and 
therefore full synchronization can never be obtained \cite{Boc06,Sen16}. 
Therefore, $S_{\rm p}$ keeps almost constant (Fig.\ \ref{bifcomp}(b)) and high 
within the entire coupling interval, except for a window of banded chaos 
where it drops. }

\begin{figure}[ht]
  \centering
  \includegraphics[width=0.46\textwidth]{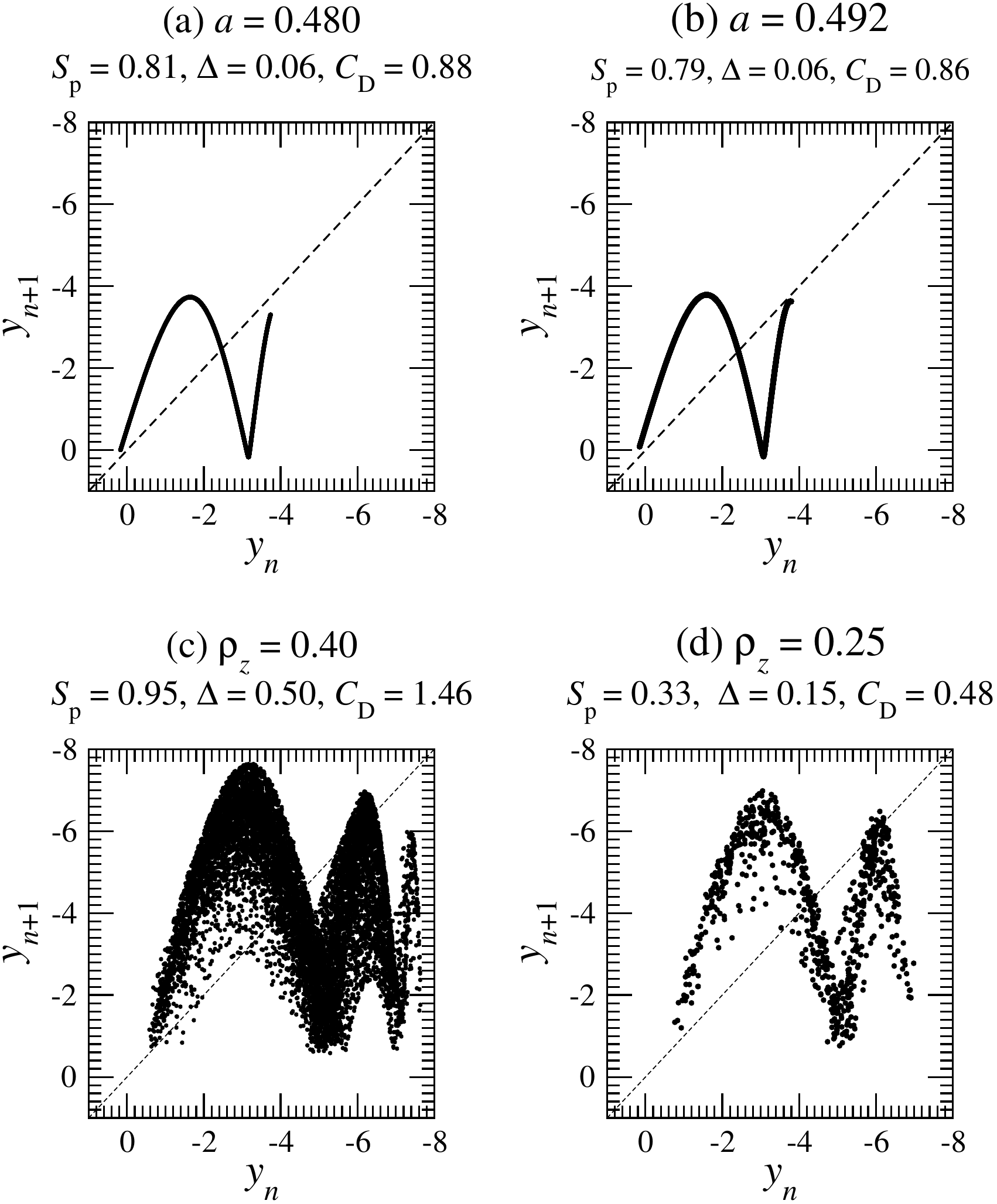} \\[-0.3cm]
  \caption{First-return maps to the Poincar\'e section Eq.~(\ref{FRMRoss}) 
produced by a dyad of R\"ossler systems when they are uncoupled (a, b) and for 
the two coupling values (c, d) marked with vertical dashed lines in 
Fig.~\ref{bifcomp}(b).}
  \label{rosmapfun}
\end{figure}

{
The first-return map to the Poincar\'e section}
\begin{equation}
 {\cal P}_{\rm Rd} \equiv   \left\{ \left( \displaystyle y_{k,i}, z_{k,i} \right) \in \mathbb{R}^2 ~|~
    x_{k,i} = x_{k,{\rm p}}, \dot{x}_{k,i} > 0
  \right\} \label{FRMRoss}
\end{equation}
where \[ x_{k,{\rm p}} = \frac{c-\sqrt{c^2 -4 a_i b}}{2} \]
is shown in Fig.~\ref{rosmapfun} for the two oscillators when uncoupled 
[Fig.~\ref{rosmapfun}(a) and \ref{rosmapfun}(b)] and for two values of the 
coupling $\rho_z$ [Fig.~\ref{rosmapfun}(c) and \ref{rosmapfun}(d)], using 
$N_{\rm p} = 10^5$ in all cases. The relative frame $[-8;0]^2$ was used.
As expected, the relative structurality for the uncoupled maps is the same and very low. When coupled, the lack of synchronization affects $\Delta$, which slowly increases within the chosen coupling range. In other words, the dynamics does not become  less predictable but more difficult to describe, as in the example in Fig.~\ref{rosmapfun} (c). 
For $0.225 < \rho_z < 0.308$, the two R\"ossler systems produce very limited banded chaos (or intermittency) but they are not synchronized. This can be appreciated in Fig. \ref{rosmapfun}(d) for a coupling within this region: $\Delta$ still remains well above the value for the single uncoupled units, signaling that the dyad is not synchronized. 

\subsection{Delay-differential Mackey-Glass equation}

Finally, let us now consider the Mackey-Glass model whose attractor has an 
embedding dimension which scales with the delay \cite{Far82}. 
{The Mackey-Glass equation is a nonlinear delay differential equation 
\cite{Mac77} which can be written in the form}
\begin{equation}
  \label{redMG}
  \dot{x} = \mu \frac{x_\tau}{1 +x_\tau^n} - x \, .
\end{equation}
{
Simulations were performed with $\tau \in [1.5;8]$ and the corresponding 
complexity markers are shown in Fig.~\ref{bifcomp}(c). }

\begin{figure}[ht]
  \centering
  \includegraphics[width=0.45\textwidth]{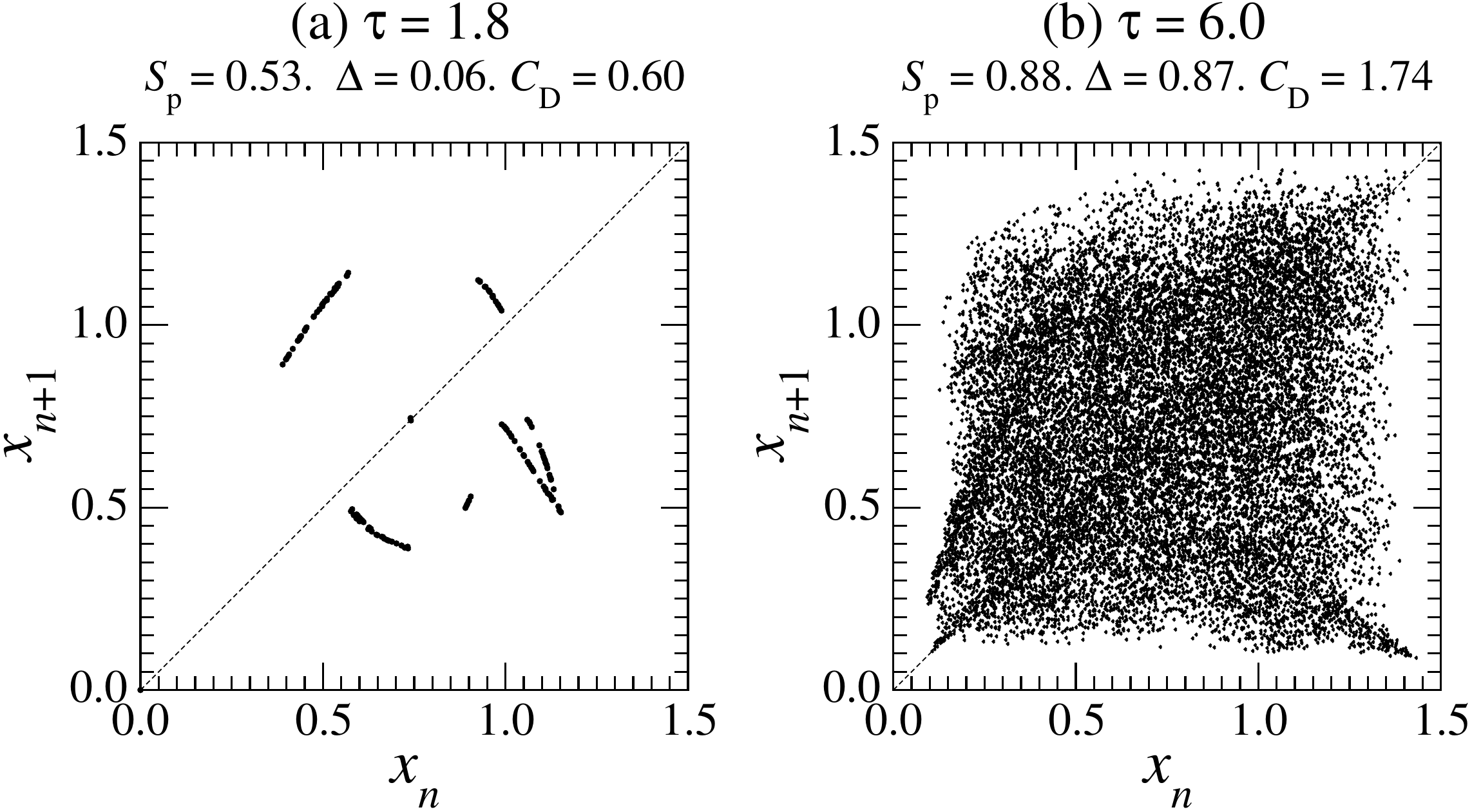} \\[-0.3cm]
  \caption{Dynamical regimes produced by the Mackey-Glass equation
(\ref{redMG}) for two delay values.  {The corresponding $\tau$ values are marked with vertical dashed lines in Fig.~\ref{bifcomp}(c).}  Other parameter values: $\mu = 2$, and
$n = 10$. }
  \label{mkgmaps}
\end{figure}

{
The Poincar\'e section was defined as:}
\begin{equation}
  {\cal P}_{\rm MG} \equiv 
  \left\{ \displaystyle x_n ~|~ \dot{x}_n = 0, \ddot{x}_n > 0 \right\} \, . 
\end{equation}
The relative frame was defined as $[0.06;1.44]^2$ and in all cases 
$N_{\rm p} = 20,000$. Two typical first-return maps are shown in Fig.\ 
\ref{mkgmaps}, for a low-dimensional case [Fig.\ \ref{mkgmaps}(a)] and a very 
high dimensionality [Fig.\ \ref{mkgmaps}(b)]. 

It can be observed [Fig.~\ref{bifcomp}(c)] that, after an initial 
period-doubling cascade (not shown), $S_{\rm p}$ goes up to $0.5$ while 
$\Delta$ is very low, in agreement with the creation of a chaotic 
attractor with a nearly 1D first-return map [Fig.~\ref{mkgmaps}(a)]. Beyond 
$\tau=2$, the dynamics becomes more difficult to describe {as shown in 
Fig.~\ref{mkgmaps}(b)} for $\tau=6$, with a much faster growth of the relative 
structurality which converges to one.   

\section{Data from the real world}
\label{data}

\subsection{Electrochemical dissolutions}

{
We use our method to characterize the complex dynamics of experimental data 
coming from electrochemical dissolutions of copper and iron rods as described 
in Refs.~\cite{Let95c,Let18c}. The experimental setup consisted of a rotating 
disk electrode in acid solution which had a copper or an iron rod embedded in a 
Teflon cylinder. A cylindrical platinum net band (much larger than the disc) 
was put around the  disk as a counter electrode to get uniform potential and 
current distributions.}

\begin{figure}[ht]
  \begin{center}
    \includegraphics[width=0.48\textwidth]{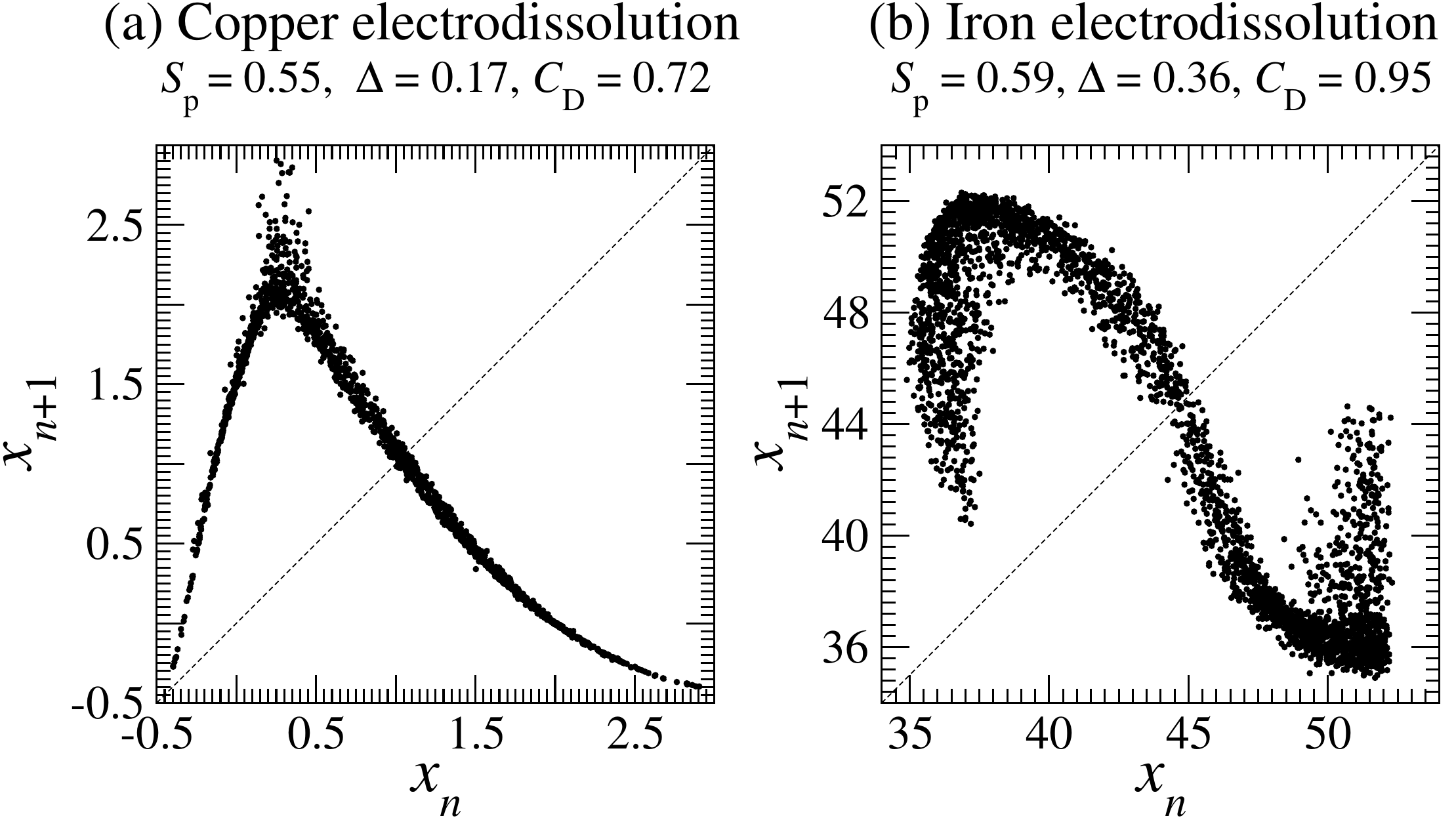} \\[-0.3cm]
    \caption{First-return maps built with the experimental data collected from 
two electrochemical dissolutions of (a) copper ($N_{\rm p} = 1,708$) and (b) 
iron ($N_{\rm p} = 3,180$) electrodes.}
   \label{expatmaps}
  \end{center}
\end{figure}

{
The dynamics is investigated from the measurements of the current $X = I$ 
flowing in the anode. A state space is reconstructed using the successive 
derivatives $Y = \dot{I}$ and $Z = \ddot{I}$. In Fig.\ \ref{expatmaps} are 
shown the first-return maps to the Poincar\'e section of the two experiments, 
defined as
}
\begin{equation}
  {\cal P}_{\rm ED} \equiv 
  \left\{ \displaystyle (X_n, Z_n) \in \mathbb{R}^2 ~|~ Y_n = 0, \dot{Y}_n >0 
  \right\}
\end{equation}
{
For the copper electrodissolution (Fig.\ \ref{expatmaps}(a)), the first-return 
map is a smooth unimodal map corresponding to a chaotic regime compatible with 
a period-doubling cascade as a route to chaos. The copper attractor was shown 
to be topologically equivalent to the R\"ossler attractor \cite{Let95c} by 
extracting the unstable periodic orbits from the experimental data and finding 
that this chaotic regime was characterized by the kneading sequence (100110) as 
observed in the R\"ossler attractor for $a = 0.424$, $b = 2$, and $c=4$ 
\cite{Dut95}. Typically, the permutation entropy should be the same in these 
two cases, and the renormalized structurality only slightly greater for the 
experimental data due to the noise contamination. Complexity values for the 
copper electrodissolution and the equivalent R\"ossler system are reported in 
Table~\ref{expres} for $N_{\rm p}=1,708$, which is the  {sample size of the} available experimental 
data. In the case of the R\"ossler system, we also computed the markers for 
$N_{\rm p} = 10^5$, to check the possible effect of the reduced number 
of experimental points, and there are no significant differences between the 
complexity markers of the two deterministic time series. Therefore, we conclude 
that the reduced number of points in the experimental data is not having a 
strong influence. However, as expected, the renormalized structurality is 
clearly larger for the experimental data than for the R\"ossler attractor, 
mainly coming from noise contamination. Noise thus rends less describable the 
dynamics, and the structurality allows to quantify this difference that was 
undetectable by $S_{\rm p}$. 
}

\begin{table}[ht]
  \centering
  \caption{Permutation entropy $S_{\rm p}$, renormalized structurality $\Delta$, and 
dynamical complexity $C_{\rm D}$ for the eletrochemical dissolutions of copper and iron and for the equivalent R\"ossler system to the copper 
electrochemical dissolution.}
  \label{expres}
  \begin{tabular}{lcccc}
    \\[-0.3cm]
    \hline \hline
    \\[-0.3cm]
    & $N_{\rm p}$ & $S_{\rm p}$ & $\Delta$ & $C_{\rm D}$ \\[0.1cm]
    \hline
    \\[-0.3cm]
    Copper & 1,708 & 0.57 & 0.11 & 0.69 \\[0.1cm]
    R\"ossler & 1,708 & 0.58 & 0.04 & 0.62 \\
              & $10^5$ & 0.57 & 0.03 & 0.60 \\[0.1cm]
    \hline
    \\[-0.3cm]
    Iron & 3,180 & 0.62 & 0.30 & 0.93 \\[0.1cm]
    \hline \hline
  \end{tabular}
\end{table} 

{
The dynamics of the iron electrochemical dissolution (Fig.~\ref{expatmaps}(b)) 
is known to be more complex than the copper one \cite{Let18c} with a 
correlation dimension about $2.4$ \cite{Wan91} while it is about $2.0$ for the 
copper. The difficulty in this dynamics results from the thickness of the 
first-return map and the shape of its rightmost part where, most likely, 
different branches should have been distinguished since, as suggested in Ref.
\cite{Let18c}, a five strip template could be underlying the iron dynamics. The 
larger complexity, as revealed by the higher value of $C_{\rm D}$, cannot be 
explained by an increase of $S_{\rm p}$, whose value is equivalent to the one 
obtained in the copper experiment. However, the renormalized structurality 
$\Delta_{\rm iron}$ is twice $\Delta_{\rm copper}$, which correctly returns 
that the iron dynamics is less describable than the copper one 
(Table~\ref{expres}), in agreement with other analyses \cite{Let18c,Wan91}.
}

\subsection{Cardiac variability}

The applicability of our measure is  demonstrated here by using it as a 
discriminating biomarker for different cardiac dynamics and associated 
pathologies recorded through electrocardiograms (ECG). Most of the data are
freely available at the {\sc PhysioNet} website \cite{Gol00}, a research 
resource for complex physiologic signals. In an ECG, each beat is associated 
with a  QRS complex corresponding to the ventricular systole. The letter ``R'' 
designates the associated large positive oscillation. The duration between two 
successive oscillations R are designated by the RR interval, and roughly 
corresponds to the duration of a cardiac beat. In the present case, heart rate 
variability is here investigated from the 
\[ \Delta \mbox{RR}_n = \mbox{RR}_n - \mbox{RR}_{n-1} \]
computed from two consecutive RR intervals.  

\begin{figure}[htbp]
  \centering
     \includegraphics[width=0.48\textwidth]{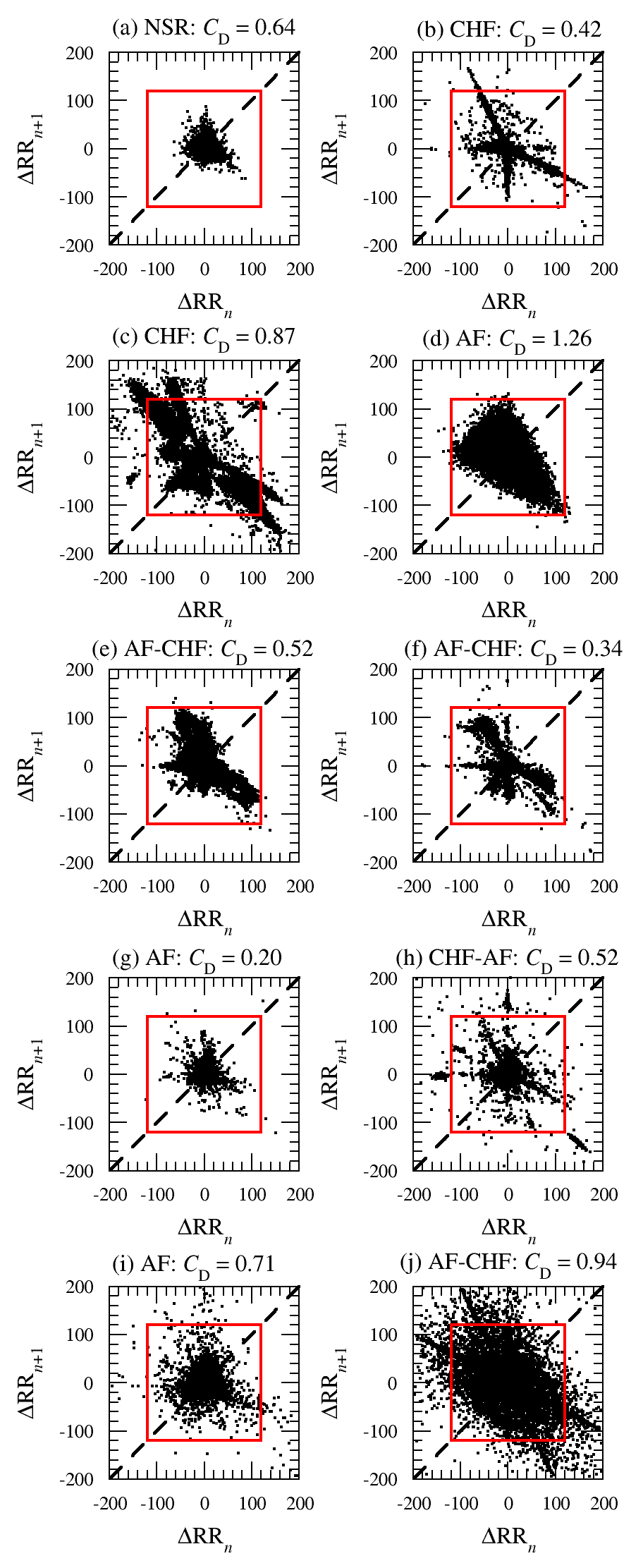} \\[-0.3cm]
  \caption{Examples of first-return maps built from the $\Delta$RR$_n$ of
patients diagnosed with normal sinus rhythm (a), congestive heart failure 
(b, c), atrial fibrillation (d, f), and preterm infants (g, j). While 
conditions reflected in (a  f) are diagnosed cases, preterm infants were 
classified as AF and CHF-AF following our expertise.}
  \label{rrmaps}
\end{figure}

According to a previous study \cite{Fre15}, a first-return map based on the 
$\Delta \mbox{RR}_n$ allows an efficient discrimination between different 
cardiac dynamics, using as markers the permutation entropy and an asymmetry 
coefficient measuring the occurrences of null, positive and negative 
$\Delta \mbox{RR}_n$. Here we replace the asymmetry coefficient by our relative
structurality $\Delta$, which is a more general marker by definition. { As a control group, we analyzed 18 long-term ECGs recorded in healthy subjects \cite{Gol00}. In Fig.~\ref{rrmaps}(a) we plot a typical first-return map recorded from one individual in this group. } 

To better discriminate the often subtle differences between patients 
with  various cardiac dynamics, we have to define a common reference frame to 
compute the structurality. We choose it to keep at least 85\% of the data 
points for every patient, that is, $\mid \Delta \mbox{RR}_n \mid \, < 120$~ms. 
This bound for the heart rate variability is large enough to take into account 
cardiac pathology as severe as atrial fibrillation \cite{All96}, a disorganized 
activity within the atria which causes the contraction of the ventricles at 
seemingly-random intervals and which is associated with a strong stochastic 
component \cite{Gol67,Mei68,Boo70}. Thus, the retained relative frame, marked 
with a dotted red line in Figs.~\ref{rrmaps}, bounds a typical atrial 
fibrillation (AF) as the one shown in Fig.~\ref{rrmaps}(d). Permutation are 
allowed when $\mid \Delta \mbox{RR}_{n+1} - \Delta \mbox{RR}_n \mid > 5$~ms.

Figure~\ref{rrmaps} shows the first-return maps of the heart rate variability 
of subjects suffering from different cardiac pathologies and ages. A group of 
15 patients with congestive heart failure (CHF) was investigated in 
Ref.~\cite{Bai86}. Two typical examples are shown in Figs.~\ref{rrmaps}(b)
and \ref{rrmaps}(c). Patient 9 has only isolated extrasystoles 
[Fig.~\ref{rrmaps}(b)] and patient 2 [Fig.~\ref{rrmaps}(c)] has bursts of 
extrasystoles as evidenced by the additional (thick) segments. Another group of 
15 long-term ECGs was recorded with subjects suffering from paroxysmal or 
persistent AF. Four different examples are shown in 
Figs.~\ref{rrmaps}(d)-\ref{rrmaps}(f). Patient 15 presents fully developed AF 
with a typical triangular shape [Fig.~\ref{rrmaps}(d)]. AF and CHF can promote 
each other and can also be found combined in the same patient 
\cite{Bur08,Pell15}, as in patient number 5 [(Fig.~\ref{rrmaps}(e)] whose 
first-return map displays a central triangular cloud like in AF around which 
there are some segments typically associated with CHF. We also investigated 
ECGs recorded from ten preterm infants \cite{Gee17} 
[Figs.~\ref{rrmaps}(g)-\ref{rrmaps}(j)], and from infants with sudden-death 
risk (not shown) \cite{YacPhD}. This data set was also analyzed in a previous 
study where the first-return maps of $\Delta$RR$_n$ were introduced 
\cite{Fre15}. AF can occur with intermittencies \cite{Cam12}.

\begin{figure}
  \centering
  \includegraphics[width=0.4\textwidth]{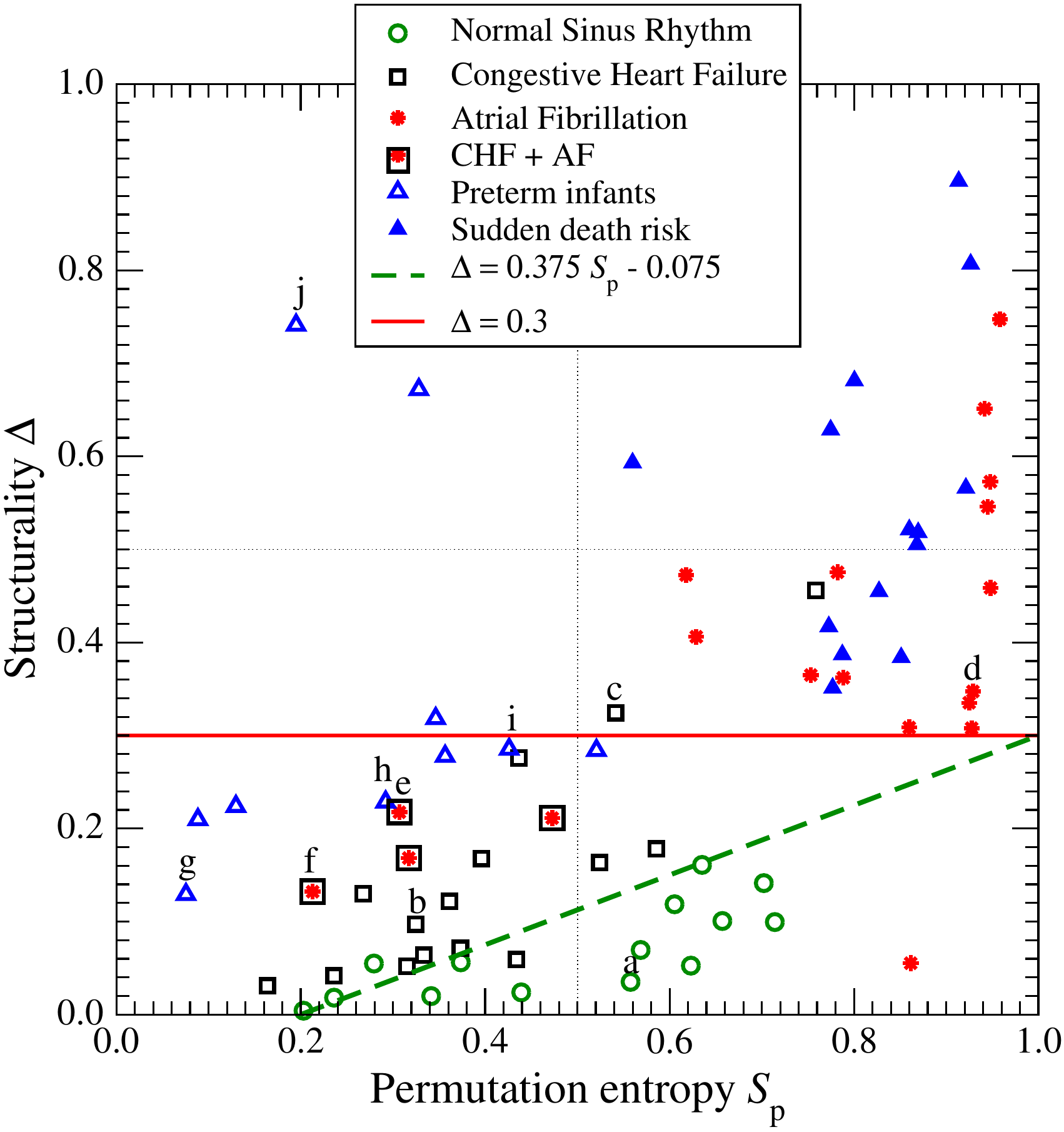} \\[-0.3cm]
  \caption{Permutation entropy $S_{\rm p}$ vs relative structurality 
$\Delta$ computed for cardiac dynamics data from various groups of patients. 
Each symbol corresponds to a given patient. See the legend to read the type of 
diagnosed cardiac dynamics. The equations of the continuous and dashed lines 
are also given in the legend. { Symbols corresponding to the 
patients reported in Fig.~\ref{rrmaps} are here highlighted.}}
  \label{HRV-SDmap}
\end{figure}

In  Fig.~\ref{HRV-SDmap} we present a general view of the analysis in the $S_{\rm p}$-$\Delta$ plane, where it can be observed how the different populations typically organize around specific areas.  Healthy subjects  are characterized by small structurality values bounded by  $\Delta < 0.375 S_{\rm p} -0.075$ and $0.2 < S_{\rm p} < 0.7$, which is reflecting that a normal sinus rhythm is irregular due to the variability in the breathing rate and activity of the autonomic nervous system. However, abnormal rhythms like CHF and AF have also features clearly distinguishable with our dynamical complexity measure. Patients with CHF 
exhibit well marked segments in the \rm{RR} sequence associated with ectopic 
beats characterized by $(S_{\rm p},\Delta)$ pairs above the straight line 
$\Delta > 0.375 S_{\rm p} -0.075$ and $S_{\rm p} < 0.7$. AF induces relative structurality $\Delta >0.3$ and in most cases has an entropy $S_{\rm p} > 0.6$ depending on how sustained is the AF. 
Some intermediate cases occur when patients present AF combined with ectopic beats. Finally, preterm infants distribute above the CHF patients but 
with slightly higher $\Delta$ values, while infants diagnosed with sudden 
death risk have, almost all of them, $S_{\rm p} > 0.6$ and $\Delta > 0.3$, 
compatible with an AF case, the exceptions corresponding to cases developing a 
mixture of CHF and AF or paroxysmal atrial fibrillation. 

To assess the significance in the differences between the various groups of 
patients, we applied a $T$-test on the permutation entropy $S_{\rm p}$, the 
relative structurality $\Delta$, and the dynamical complexity $C_{\rm D}$, 
respectively. Results are reported in Tables \ref{SpD} and \ref{CD}. We 
retained 
five different groups of subjects for our statistical analysis: subjects with 
normal sinus rhythm (NSR), congestive heart failure (CHF), atrial fibrillation 
(AF), preterm newborns, and infants with sudden death risk (SDR). Permutation 
entropy $S_{\rm p}$ is significantly different among the five groups, but is 
not significantly different between NSR and CHF, as well as between SDR and AF 
(lower half of Table \ref{SpD}). The relative structurality $\Delta$ by itself 
allows to discriminate the five groups except preterm newborns from AF (upper 
half of Table \ref{SpD}). As a result of combining both markers, all the groups 
can be discriminated. Nevertheless, if the dynamical complexity is computed, it 
is not possible to distinguish NSF from CHF and preterm infants, CHF from 
preterm infants, and AF from SDR. These features result from some balance which 
can occur between the entropy $S_{\rm p}$ and the structurality $\Delta$. An 
accurate analysis is required in the $S_{\rm p}$-$\Delta$ plane to correctly 
characterize all the different groups.

\begin{table}[ht]
  \centering
  \caption{T-test to assess the significance in the difference of the permutation entropy $S_{\rm p}$ and the 
  relative structurality $\Delta$ for the various groups of patients.}
  \label{SpD}
  \begin{tabular}{cccccccc}
    \\[-0.3cm]
    \hline \hline

    \\[-0.3cm]
    & & NSR & CHF & AF & Preterm & SDR & \\[0.1cm]
    \hline
    \\[-0.3cm]
                & NSR     & --- & *   & *** & *** & *** & $\Delta$ \\
                & CHF     & NS  & --- & *** & **  & *** & \\
                & AF      & *** & *** & --- & NS  & *   & \\
                & Preterm & **  & *   & *** & --- & **  & \\
    $S_{\rm p}$ & SDR     & *** & *** & NS  & *** & --- & \\[0.1cm]
    \hline \hline
    \\[-0.3cm]
  \multicolumn{8}{p{0.45\textwidth}}{NSR $\equiv$ normal sinus rhythm, CHF
$\equiv$ congestive heart failure, AF $\equiv$ atrial fibrillation, SDR
$\equiv$ infant with sudden death risk, NS $\equiv$ Non-Significant,
$* \equiv p < 0.05$, $** \equiv p < 5 \cdot 10^{-3}$,
$*** \equiv p < 5 \cdot 10^{-4}$.}
  \end{tabular}
\end{table}

These results are partly recovered by using the sole dynamical 
complexity $C_{\rm D}$ as reported in Table \ref{CD}. {It clearly 
distinguishes NSR, CHF, and preterm from SDR and AF.} These markers, whose 
computation cost is very reduced, allow therefore to correctly assess the 
complexity underlying the heart variability.

\begin{table}[ht]
  \centering
  \caption{$T$-test to assess the significance in the difference of the 
dynamical complexity $C_{\rm D}$  for the various groups of patients. Same 
terminology as described in the caption of Table \ref{SpD}.}
  \label{CD}
  \begin{tabular}{cccccccc}
    \\[-0.3cm]
    \hline \hline
    \\[-0.3cm]
    & & NSR & CHF & AF & Preterm & SDR & \\[0.1cm]
    \hline
    \\[-0.3cm]
                & NSR     & --- &     &     &     &     & \\
                & CHF     & NS  & --- &     &     &     & \\
                & AF      & *** & *** & --- &     &     & \\
                & Preterm & NS  & NS  & *** & --- &     & \\
                & SDR     & *** & *** & NS  & *** & --- & \\[0.1cm]
    \hline \hline
  \end{tabular}
\end{table}

\section{Conclusions}

We have shown that to distinguish between organized and disorganized 
complexity, a marker capable of detecting a structured dynamics is needed 
independently of its degree of predictability. We propose a complexity measure 
combining these two notions of unpredictability, assessed with a permutation 
entropy, and that of structurality which quantifies, in a Poincar\'e section, 
how the structure underlying a dynamics can be described. The boundary 
conditions of our complexity measure are such that it vanishes for regular 
motion and whose upper limit reflects the disorganized dynamics of a stochastic 
signal, neither predictable nor easily describable in the Poincar\'e map. It 
thus provides a powerful measure for characterizing any stationary dynamics 
produced by either a map or a flow (as long as the time series can be 
investigated in a Poincar\'e plane), dissipative or conservative. It should be 
noted that our measure is not extensive and as such the upper limit can 
correspond to dynamics that can greatly differ in dimension. This problem can 
be addressed by using a Poincar\'e section whose dimension is greater than two 
but is out of the scope of the present work. As an illustration of its 
classifying power, we evidenced that this complexity measure can discriminate 
among various groups of common cardiac diseases from the sole measure of an ECG 
and we are confident that it will be useful for a reliable characterization of 
a large variety of real-world dynamics.

\bigskip

\acknowledgments ISN and IL acknowledge
partial support from the Ministerio de Econom\'ia y Competitividad of Spain 
under Project No. FIS2017-84151-P.

\appendix
\section{Robustness of the dynamical complexity }
\label{detail}

\begin{figure}[t]
  \centering
  \includegraphics[width=0.48\textwidth]{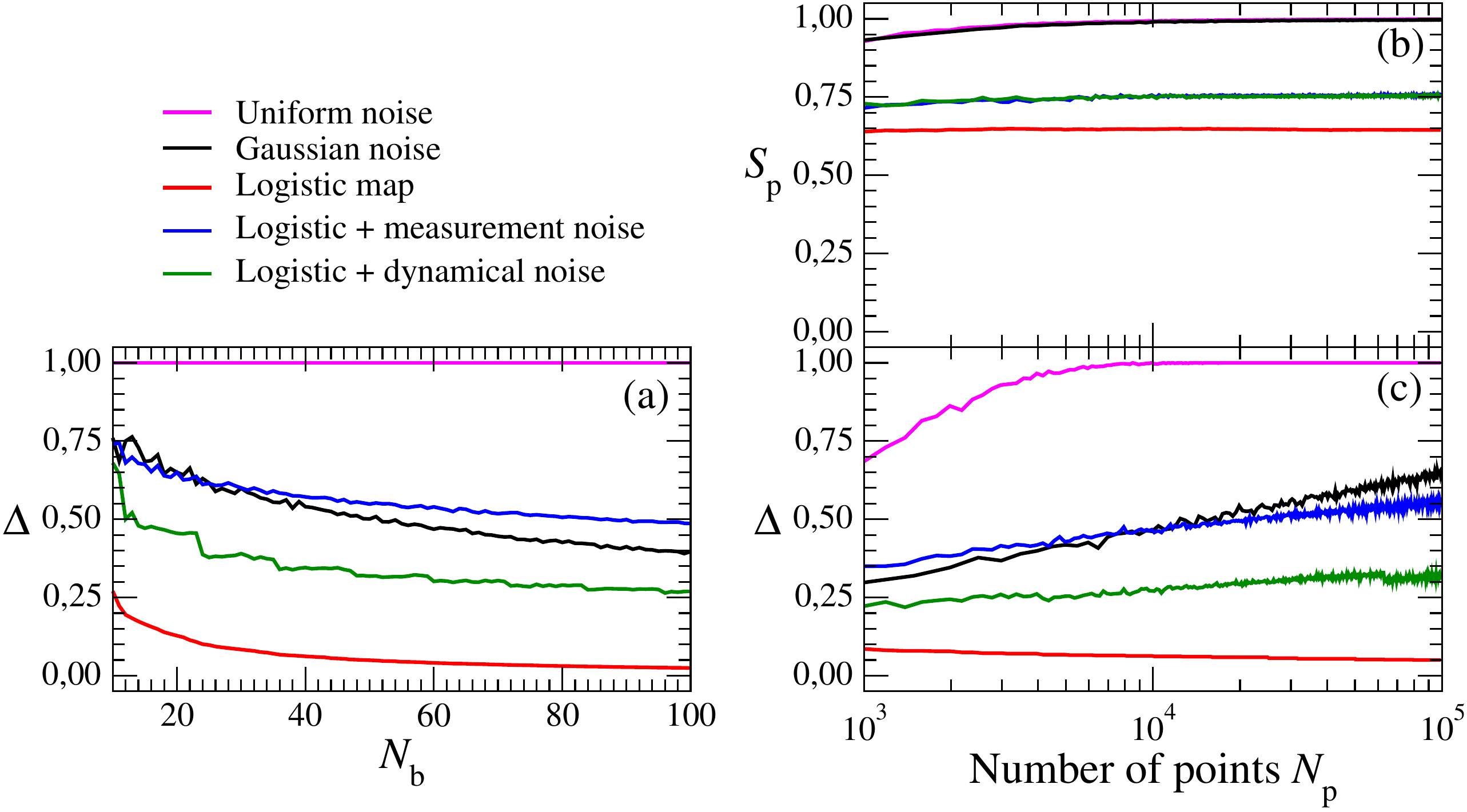}  \\[-0.3cm]
  \caption{(a) Dependence of the renormalized structurality $\Delta$ on 
$N_{\rm b}$ (square root of the total number of boxes of the pixelation) for 
the deterministic logistic map for $\mu=3.99$ (red curve) and for the logistic 
map with Gaussian observational noise of standard deviation (std) 
$0.05$ ($B=0.2$) (blue curve), and with additive Gaussian dynamical noise of 
std $0.025$ ($B=0.1$) (green curve). As a reference, uniform noise (magenta) 
and Gaussian noise of std $0.25$ (black) are shown. In all cases, 
$N_{\rm p}=10^5$ are used. Dependence of the permutation entropy $S_{\rm p}$ 
(b) and structurality $\Delta$ (c) on the number of points $N_{\rm p}$ for the 
same cases indicated in the legend in the left panel. In all cases, 
$N_{\rm b} \sim 10 \log_{10}N_{\rm p}$.}  
  \label{depNdat}
\end{figure}

{
In the following, we will provide an extensive analysis of the factors involved 
in the definition of the structurality $\Delta$ and how it behaves under the 
presence of different noise sources. The logistic map $x_n$ in 
Eq.~(\ref{logistic}) for $\mu = 3.99$ was studied for different pixelation 
settings and lengths of the time series. In Fig.~\ref{depNdat}(left panel) we 
varied the number of boxes $N_{\rm b}^2$ in which is divided the 
renormalized frame of the first-return map for a time series of length 
$N_{\rm p}=10^5$. We observe (red curve) how the structurality converges to a 
very low value, as it should be for a well structured dynamics, when the number 
of boxes $N_{\rm b}>50$ (along one dimension). This points out to a possible 
parametrization of the number of boxes as a function of the number of points as 
$N_{\rm b} \sim 10 \log_{10}N_{\rm p}$, as specified in the main text. This 
choice allows us to reduce the dependency of the results on the number of 
points $N_{\rm p}$ as it ensures that boxes are visited with a significant 
mean probability even when a small number of points ($N_{\rm p}<10^4$) is 
available. At the same time, the logarithmic dependence avoids the redundancy 
that could yield to underestimate $\Delta$. This is shown in the right column 
of the same Fig.~\ref{depNdat}, where both the permutation entropy $S_{\rm p}$ 
and the renormalized structurality  $\Delta$ of the deterministic logistic map 
are almost independent of $N_{\rm p}$ when considering time series of lengths 
ranging from $N_{\rm p}=10^3$ up to $10^5$. Notice that the permutation 
entropy keeps a constant value when $N_{\rm p} > 10^4$.
}

Regarding the effect of fluctuations present in the time series, we added a Gaussian noise $\xi_n$  (of zero mean and variance $0.0625$) to the deterministic time series simulating both observational ($q_n = x_n + B\xi_n$, blue curves) and dynamical ($x_{n+1} = \mu x_n ( 1-x_n)+ B\xi_n$, green curves) stochastic sources, with $B$  controlling the noise intensity.  Again, Fig.~\ref{depNdat} shows that the structurality obviously increases with respect to the pure deterministic case (red curve) but it is almost independent on the number of boxes as long $N_{\rm b}>50$. For comparison purposes, we also added the structurality of two pure stochastic processes, exhibiting intermediate values for the case of a Gaussian noise (black curve) and saturating at one for uniform noise (magenta curve) as it should be since the uniform noise is uniformly distributed in the renormalized frame. Looking at the right column of  Fig.~\ref{depNdat}, it is worth noting how the structurality is able to distinguish the logistic map with added observational and dynamical noises (blue and green curves) while the permutation entropy is not.  Another remark is that while the permutation entropy is maximum 
for both uniform and Gaussian noise (they have the same degree of 
unpredictability), the structurality (right column-bottom panel) only saturates 
for the uniform noise and linearly increases with the number of points in the case of the Gaussian noise. A final observation in this study, it is that it is safe to consider for both $S_{\rm p}$ and $\Delta$, $N_{\rm p} > 10^4$,  in agreement with what was already obtained with a Shannon entropy computed from recurrence plots \cite{Let06b}. 

\begin{figure}[t]
    \centering
    \includegraphics[width=0.44\textwidth]{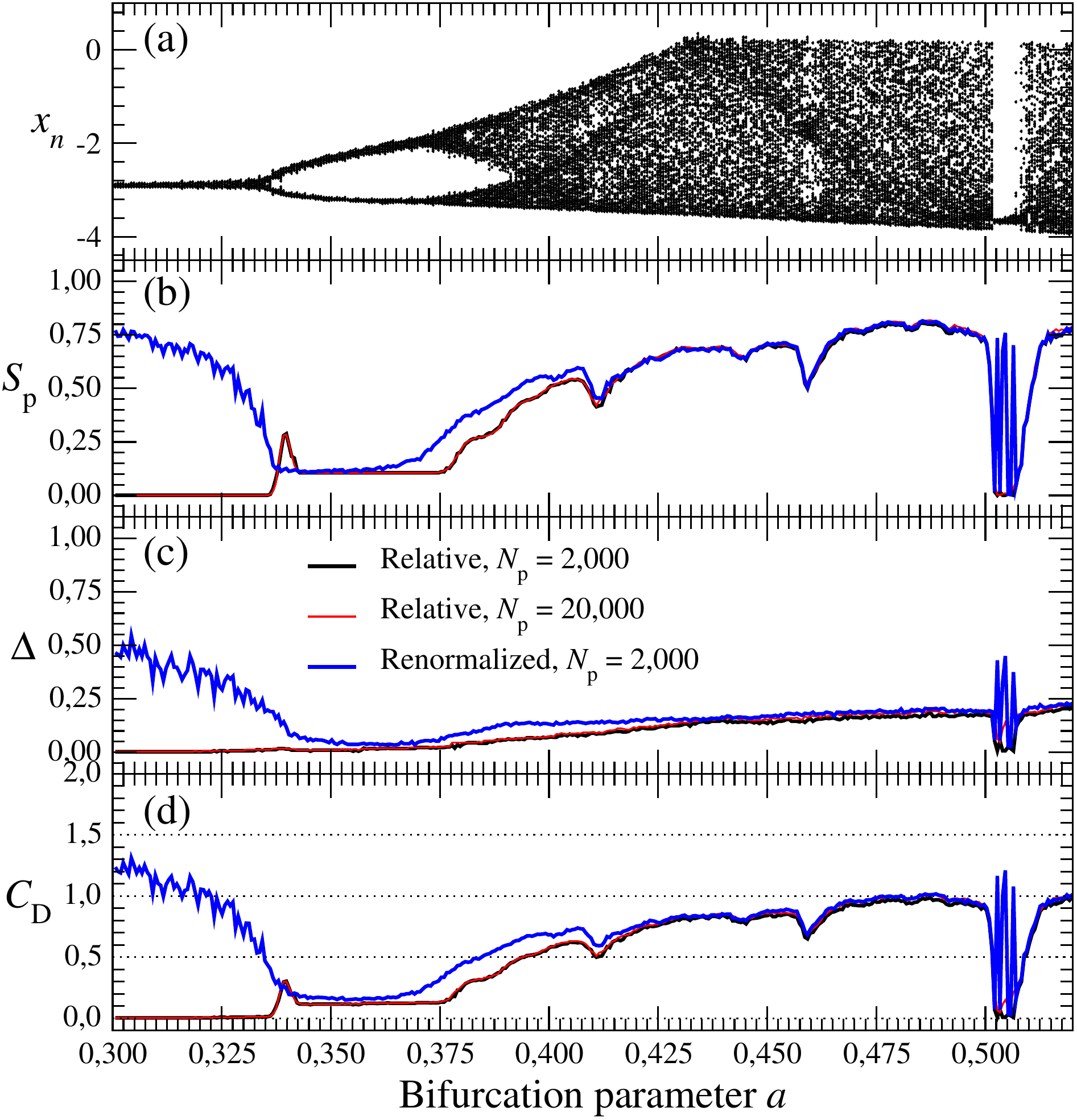}  \\[-0.2cm]
  \caption{Dynamics of the R\"ossler system with added Gaussian dynamical noise of std $0.005$ ($B=0.02$) for varying bifurcation parameter $a$. (a) Bifurcation diagram, (b) permutation entropy $S_{\rm p}$, (c) structurality $\Delta$, and (d) dynamical complexity $C_{\rm D}$. From (b) to (c), each curve denotes the corresponding marker computed using relative ($N_{\rm p}=2,000$ in black and $N_{\rm p}=20,000$ in red) and renormalized frames ($N_{\rm p}=2,000$ in blue). Other parameter values: $b=2$ and $c=4$.\label{depnoise}}
\end{figure}

Another factor affecting our structurality measure is the framing of the 
Poincar\'e section. To show how the choice of a renormalized or relative frame 
is actually contributing to $\Delta$, we computed the bifurcation diagram of 
the R\"ossler system with the parameter $a$ [Eq.~\ref{ros76} with $\rho_z=0$ 
and $a_1=a$] and influenced by a Gaussian dynamical noise of zero mean and 
standard deviation $0.005$. Since the noise contamination is quite limited, the 
dynamical complexity is mainly driven by the permutation entropy while the relative structurality (red traces in both panels) slightly increases with the parameter $a$:  the more the chaos develops, the thicker the first-return map and the larger the relative structurality. However, when considering a renormalized frame, huge discrepancies are clear between the two frames in the period-one windows. A renormalized frame is no longer able to correctly detect the deterministic part of the dynamics and over estimates the noise contribution. For instance
the entropy as well as the structurality are arbitrarily large for  $a < 0.33$ and there is a large instability in the period-1 window at $0.502 < a < 0.508$. In these period-1 windows, a relative frame provides more stable results with the three markers, $S_{\rm p}$, $\Delta$ and $C_{\rm D}$ exhibiting low values despite the noise contamination.  

\begin{figure}[b]
  \centering
  \includegraphics[width=0.49\textwidth]{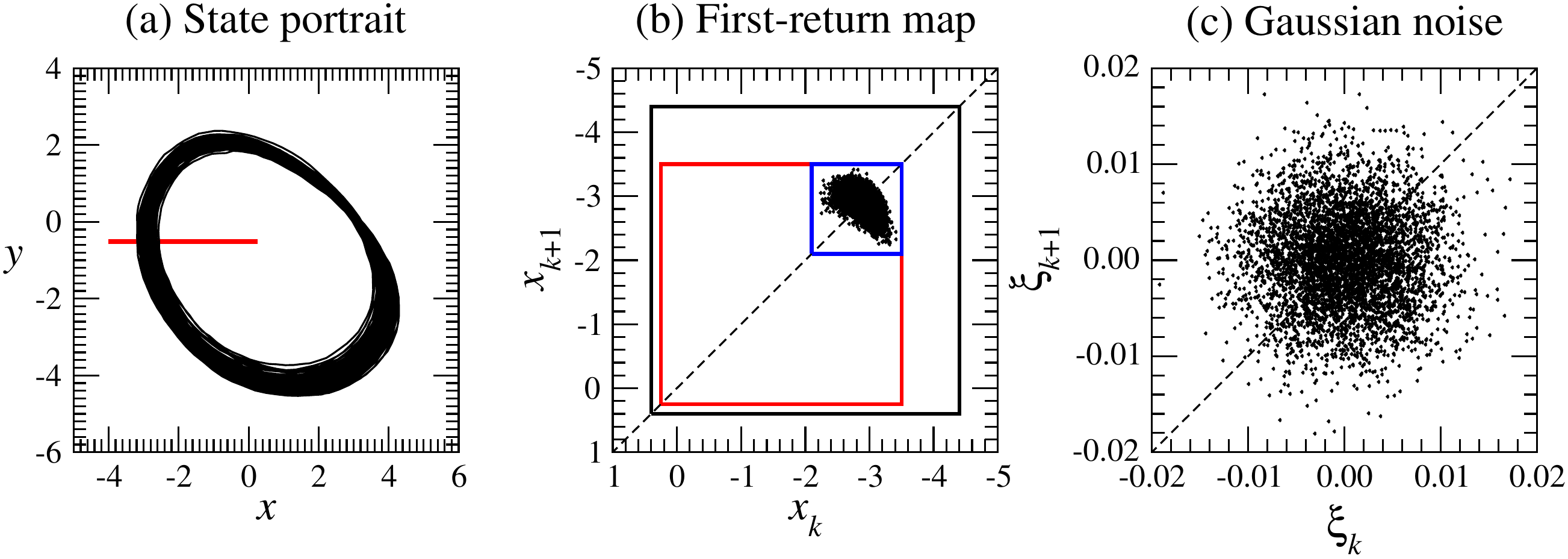}  \\[-0.2cm]
  \caption{Dependence of the complexity markers on the chosen reference frame. 
(a) State portrait of a noisy period-one limit cycle produced by one of the 
R\"ossler systems (\ref{ros76}) with a Gaussian dynamical noise of standard 
deviation $0.00375$. The center of the limit cycle is at $(0.25,-0.25)$ and the 
red line extending from $x=-3.46$ to $x=0.25$ is the projection of a Poincar\'e 
plane. (b) First-return map of the Poincar\'e section of (a) framed by the 
renormalized frame (blue square), and the relative frames of the largest 
chaotic attractor in the bifurcation diagram shown in Fig.~\ref{depnoise}(a) 
(black square) and of the period-one limit cycle (red square). (c) First-return 
map of the Gaussian noise injected in (a) and plotted in the renormalized 
frame.}
  \label{depnoise2}
\end{figure}

{
The overestimation of the noise contribution to a period-one limit cycle is 
further illustrated in Fig.\ \ref{depnoise2}. The Poincar\'e plane is plotted 
as a red line in the state portrait of Fig.\ \ref{depnoise2}(a). 
The size of the corresponding Poincar\'e section (the crossings of the limit 
cycle with the Poincar\'e plane) defines the renormalized frame 
($-3.4 < x_n < -2.15$) shown as a blue square in Fig.\ \ref{depnoise2}(b). It 
is clear that the period-one limit cycle no longer appears as a structured 
dynamics with respect to the renormalized frame due to the noise contamination. 
However, when observing the Poincar\'e section using a relative frame as the 
one defined by the size of the largest chaotic attractor of the bifurcation 
diagram (black square) or by the period-one limit cycle (red square), the noise 
contribution to the structurality is damped and the dynamics reveals to be more 
structured. Just to compare, the first-return map of the Gaussian noise 
injected in the limit cycle is shown in Fig.~\ref{depnoise2}(c) in a 
renormalized frame. The values of the three complexity markers for each one of 
the three frames shown in Fig.~\ref{depnoise2}(b) are reported in 
Table\ \ref{p1lc}. There it is clear the relevance of the frame definition, 
showing a significant difference between choosing a relative or a renormalized 
frame. 
}

{
Finally, note that to avoid overestimating the entropy, permutations are 
performed only when $| x_i - x_{i+1}| > \epsilon$ with 
$\epsilon = 5 l_{\rm b}$. This kind of noise filter is in fact related to 
uncertainties inherent to measurements and allows to reduce the typical peaks 
appearing at the bifurcations in a period doubling cascade when observational 
noise is affecting the dynamical system (see Fig.\ 2e in Ref.~\cite{Ban02}) or 
avoids to have large entropy in the case of a noisy period-one limit cycle 
(Fig.~\ref{depnoise}). 
}

\begin{table}
  \centering
  \caption{Values of the complexity markers $S_{\rm p}$, $\Delta$, and 
$C_{\rm D}$ for the the noisy period-one limit cycle shown using the three 
different frames marked in Fig. \ref{depnoise2}(b).\label{p1lc}}
  \begin{tabular}{lccc}
    \\[-0.4cm]
    \hline \hline
    \\[-0.3cm]
    Frame & $S_{\rm p}$ & $\Delta$ & $C_{\rm D}$ \\[0.1cm]
    \hline
    \\[-0.3cm]
    Relative (largest attractor)  & 0.06 & 0.04 & 0.10 \\[0.1cm]
    Relative (limit cycle radius) & 0.16 & 0.07 & 0.23 \\[0.1cm]
    Renormalized & 0.70 & 0.44 & 1.14 \\[0.1cm]
    \hline \hline
  \end{tabular}
\end{table}

%

\end{document}